\documentclass[%
preprint,
amsmath,amssymb,
aip,
jcp,    
floatfix
]{revtex4-2}

\usepackage{graphicx}
\usepackage{dcolumn}
\usepackage{bm}
\usepackage{color}
\usepackage{tabularx}
\usepackage{xcolor}
\usepackage[normalem]{ulem} 
\definecolor{zgreen}{rgb}{0.0, 0.26, 0.15}

\newcommand{\quasi}[1]{{\it quasi}}
\long\def\comment#1{}

\begin{document}
\title{Self-consistent implementation of locally scaled self-interaction-correction method}
\author{Yoh Yamamoto}
\author{Tunna Baruah}
\author{Po-Hao Chang}
\author{Selim Romero} 
\author{Rajendra R. Zope}
\email{rzope@utep.edu}
\affiliation{Department of Physics, University of Texas at El Paso, El Paso, Texas 79968}

\date{\today}

\begin{abstract}
Recently proposed local self-interaction correction (LSIC) method  [Zope, R. R. \textit{et al.}, \textit{J. Chem. Phys.} {\bf 151}, 214108 (2019)] is a one-electron self-interaction-correction (SIC) method that uses an iso-orbital indicator to apply the SIC at each point in space by scaling the exchange-correlation and Coulomb energy densities. The LSIC method is exact for the one-electron densities, also recovers the uniform electron gas limit of the uncorrected density functional approximation, and reduces to the well-known Perdew-Zunger SIC (PZSIC) method as a special case. This article presents the self-consistent implementation of the LSIC method using the ratio of  Weizs\"acker and Kohn-Sham kinetic energy densities as an iso-orbital indicator. The atomic forces as well as the forces on the Fermi-L\"owdin orbitals are also implemented for the LSIC energy functional. Results show that LSIC with the simplest local spin density functional predicts atomization energies of AE6 dataset better than some of the most widely used GGA functional (e.g. PBE) and barrier heights of BH6 database better than some of the most widely used hybrid functionals (e.g. PBE0 and B3LYP). The LSIC method [mean absolute error (MAE) of 0.008 \AA] predicts bond lengths of a small set  of molecules better than the PZSIC-LSDA (MAE 0.042 \AA) and LSDA (0.011 \AA). This work shows that accurate results can be obtained from the simplest density functional by removing the self-interaction-errors using an appropriately designed SIC method. 
\end{abstract}

\maketitle

\section{\label{sec:introduction}Introduction}
The Kohn-Sham formulation of density functional theory (DFT)\cite{PhysRev.140.A1133,levy1979universal} with suitable density functional approximations (DFA) is a powerful tool in many-body quantum physics, but its predicting capabilities are often limited by notorious self-interaction error (SIE) that arises from an incomplete cancellation of self-Hartree energy with self-exchange-correlation energy. The DFT is an exact theory, but the approximation required for its practical implementation suffers from the SIE. Many failures of DFAs have been attributed to the SIEs.\cite{https://doi.org/10.1002/wcms.1631} The self-interaction correction (SIC) methods  to remove SIE in an orbitalwise fashion have been devised long ago.\cite{lindgren1971statistical,perdew1979orbital,PhysRevA.15.2135,PhysRevB.23.5048,lundin2001novel,zunger1980self,gunnarsson1981self,PhysRevA.38.630,https://doi.org/10.1002/jcc.540120705} Besides the one-electron SIC methods that remove SIE in an orbital wise fashion, several other approaches  such as von Weizsäcker kinetic energy density-based SICs (an exchange functional by Becke and Roussel derived from Taylor expansion of exchange hole,\cite{PhysRevA.39.3761,https://doi.org/10.1002/qua.560230605} regional SIC,\cite{doi:10.1002/jcc.10279,doi:10.1063/1.4866996} and local hybrid functional\cite{jaramillo2003local,doi:10.1063/1.2795700,doi:10.1063/1.4865942}), long-range asymptotic corrections,\cite{PhysRev.99.510} Koopmans-compliant functionals,\cite{dabo2014piecewise,PhysRevB.90.075135} atomic SIC (ASIC),\cite{PhysRevB.75.045101}  multiconfiguration pair-DFT (MC-PDFT),\cite{li2014multiconfiguration} and DFT+U approach\cite{anisimov1991band,PhysRevB.52.R5467} have been proposed to remove or mitigate SIE. One of the most widely used approach to mitigate effects of SIE is to mix DFAs with the Hartree-Fock  exchange using various criteria.\cite{doi:10.1063/1.464304,iikura2001long,jaramillo2003local,janesko2021replacing} Amongst the one-electron SIC methods, the Perdew and Zunger (PZSIC)\cite{PhysRevB.23.5048} method is the most well-known method. In fact, PZSIC has become synonymous with SIC. Although not as widely used as standard gradient based DFAs, a number of researchers have adopted  PZSIC method.\cite{Harrison_1983,doi:10.1063/1.445296,Heaton_1984,heaton1980electronic,Gudmundsd_ttir_2015,JONSSON20151858,doi:10.1021/acs.jctc.6b00622,doi:10.1063/1.4752229,doi:10.1021/ct500637x,doi:10.1063/1.1897378,doi:10.1063/1.1794633,doi:10.1021/acs.jctc.6b00347,doi:10.1021/acs.jpclett.2c01359,Bylaska_2006,doi:10.1063/1.2566637,doi:10.1063/1.2387954,Jackson_2019,Ziegler2002,doi:10.1080/0026897031000094506,korzdorfer2008self,gunnarsson1981self,PhysRevA.50.2191,doi:10.1080/00268970110111788,heaton1983self,goedecker1997critical,svane1996electronic,svane2000self,rieger1995self,zope1999atomic,hamada1986self,biagini1994self,xie1999obtaining,arai1995electronic,stengel2008self,price1999application} PZSIC improves the description of stretched bonds\cite{doi:10.1063/1.2566637} and gives significant improvement over DFA for barrier heights,\cite{doi:10.1063/1.5087065} atomic anions,\cite{PhysRevB.23.5048} etc. It is, however, well known that PZSIC tends to overcorrect, particularly for the equilibrium properties, resulting in errors of opposite sign to those from semilocal functionals.\cite{vydrov2006scaling,perdew2015paradox} PZSIC predicts bond lengths that are too short\cite{goedecker1997critical,csonka1998inclusion,doi:10.1063/1.5125205,vydrov2006scaling} and  provides  a little  improvement for reaction energies.\cite{doi:10.1063/1.5129533} Early efforts in the development and application of SIC methods are reviewed in the classic article by Perdew and Zunger.\cite{PhysRevB.23.5048} Another  perspective on SIC methods is by Perdew and Pederson.\cite{Pederson_5scientific} In recent years, a unitary invariant implementation\cite{doi:10.1080/0026897031000094506} of PZSIC was proposed that uses orthgonalized Fermi orbitals (FO).\cite{Luken1982,Luken1984} The resultant Fermi-L\"owdin orbital SIC (FLOSIC) method\cite{doi:10.1063/1.4869581,doi:10.1063/1.4907592,PEDERSON2015153,PhysRevA.95.052505,PhysRevA.103.042811,doi:10.1063/5.0031341} has been used by a few research groups to study wide areas of electronic structure properties such as atomization energies,\cite{doi:10.1063/5.0031341,doi:10.1063/5.0041646,romero2021local,schwalbe2020pyflosic,doi:10.1063/1.5129533,yamamoto2020assessing,doi:10.1063/5.0004738,doi:10.1063/5.0010375} static dipole polarizability of molecules,\cite{doi:10.1063/5.0041265,akter2021well,PhysRevA.100.012505,waterpolarizability} magnetic exchange coupling,\cite{joshi2018fermi,doi:10.1021/acs.jpca.1c10354} chemical reaction energies and barriers,\cite{doi:10.1063/5.0070893,doi:10.1021/acs.jpca.0c06485,doi:10.1063/1.5087065} transition metal ions and molecules,\cite{doi:10.1063/5.0054439} ionization energy,\cite{doi:10.1002/jcc.25586,https://doi.org/10.1002/jcc.26168,FLOSICSCANpaper,doi:10.1021/acs.jctc.8b00344,doi:10.1063/1.4996498} electrostatic dipole moments,\cite{doi:10.1063/1.5125205,doi:10.1063/5.0034545} photoelectron spectra,\cite{doi:10.1063/5.0056561} interpretation of Fermi orbital descriptor,\cite{NGUYEN2021138952,https://doi.org/10.1002/jcc.26062} water and water-ion cluster,\cite{doi:10.1063/5.0041620,Sharkas11283,C9CP06106A} dissociation energies,\cite{doi:10.1021/acs.jpca.8b09940,Jackson_2019} and bond lengths of molecules.\cite{doi:10.1002/jcc.25767}

Recently, Zope and coworkers\cite{doi:10.1063/5.0010375}  introduced the locally scaled SIC  method (LSIC), which uses a pointwise iso-orbital indicator to identify the one electron self-interaction regions in many-electron system (see Section~\ref{sec:localscaling}) and to determine the magnitude of SIC in the many-electron regions. LSIC was conceptualized while pruning dense numerical grids required for SIC calculations with meta-GGA functionals\cite{FLOSICSCANpaper,yamamoto2020assessing} and was inspired by the regional SIC\cite{doi:10.1002/jcc.10279} and the functionals that use iso-orbital indicators to identify single electron regions. LSIC works well for both equilibrium properties as well as for properties that involve a stretched bond, and  it provides an improved performance with respect to standard PZSIC method for a wide range of electronic structure properties.\cite{doi:10.1063/1.5129533,waterpolarizability,doi:10.1063/5.0041265,doi:10.1063/5.0041646,akter2021well} 

Early LSIC calculations\cite{doi:10.1063/1.5129533,romero2021local,doi:10.1021/acs.jpca.1c10354,doi:10.1063/5.0070893} made use of self-consistent PZSIC densities obtained from the  PZSIC method with Fermi-L\"owdin orbitals (FLO) as localized orbitals.\cite{doi:10.1063/1.4869581} The perturbative LSIC method using the simplest DFA like LSDA predicted several properties more accurately than those obtained by widely used Perdew-Burke-Ernzerhof (PBE) generalized-gradient-approximation.\cite{PhysRevLett.77.3865,PhysRevLett.78.1396} It simultaneously provides good estimates of atomization energies of AE6  data set of molecules\cite{doi:10.1021/jp035287b} (better than the PBE-GGA) and the barrier heights of BH6 database  molecules (better than popular hybrids like B3LYP, PBE0, etc) which is difficult for the most DFAs. The barrier heights of BH6 database were predicted within the chemical accuracy. The barrier heights of more diverse BH76 database are predicted within a few kcal/mol.\cite{doi:10.1063/5.0070893} Likewise, the dissociation curve of positively charged helium dimer using LSIC method was qualitatively in line with the CCSD(T) dissociation curve.\cite{doi:10.1063/1.5129533} Self-consistent implementation of the LSIC method at the time wasn't possible, as our code lacked the ability to compute the Coulomb potential due to the charge density scaled by iso-orbital indicator which is needed when  variation of LSIC  self-Hartree energy term is taken. A simplified {\it quasi}-self-consistent (quasi-SCF) scheme was instead employed in the LSIC studies of electron density-related properties such as dipole moments  and polarizabilities.\cite{waterpolarizability,doi:10.1063/5.0041265,akter2021well} The quasi-SCF approach ignores the variation of the local scaling factor and is  valid when the variation of the local scaling factor is negligible. To gauge the full potential of LSIC method, a fully self-consistent  implementation is needed. In this work we present the outline of the scheme for self-consistent LSIC method and describe its implementation using the Fermi-L\"owdin orbitals in the FLOSIC code.\cite{FLOSICcode,FLOSICcodep} We subsequently assess the performance of self-consistent LSIC by computing atomic total energies, atomization energies, reaction barrier heights, bond distances, and the highest occupied molecular orbital (HOMO) eigenvalues. These results are compared against accurate reference values along with previous one-shot and quasi-SCF results of LSIC. The results of this work show that self-consistent LSIC  method,  similar to one-shot LSIC, performs well for the properties studied in this work.

\section{\label{sec:theory}Theory and Computational Details}
Both PZSIC and LSIC are one-electron SIC methods which remove SIE in an orbital-by-orbital manner. We first briefly outline the PZSIC method\cite{PhysRevB.23.5048} and then describe the LSIC method in the next subsection. The PZSIC  total energy is given by 
\begin{equation}\label{eq:pzsic}
  E^{PZSIC} = E^{DFA}[\rho_\uparrow,\rho_\downarrow]- \sum_{i\sigma}^{occ}  (U[\rho_{i\sigma}]+E_{XC}^{DFA}[\rho_{i\sigma},0]).
\end{equation}
Here, the self-Coulomb $U[\rho_{i\sigma}]$ and self-exchange-correlation energies $E_{XC}^{DFA}[\rho_{i\sigma},0]$ are subtracted from DFA energy for each occupied orbital. PZSIC is exact for all one-electron density and nearly many-electron SIE free.\cite{doi:10.1063/1.2566637} The size extensivity requirement necessitates  use of local orbitals in the PZSIC method. In this work, we use FLOs as local orbitals.\cite{Luken1982,Luken1984} In particular, we used the FLOSIC approach of Pederson \textit{et al.}\cite{doi:10.1063/1.4869581,PEDERSON2015153} FLOSIC is a FLO implementation of PZSIC where the SIC energies are obtained with FLOs. FLOs are localized orbitals and can be obtained from KS orbitals using what is called Fermi orbital descriptor (FOD) positions. Using FODs, FOs are obtained from KS orbitals as
\begin{equation}\label{eq:fods}
  F_{i \sigma}(\mathbf{r}) = \sum_j^M \frac {\psi_{j \sigma}^*(\mathbf{a}_{i \sigma}) \psi_{j \sigma}(\mathbf{r}) } {\sqrt{\rho_{\sigma}(\mathbf{a}_{i \sigma})}} \,.
\end{equation}
FOs are then orthogonalized with the L\"owdin's orthogonalization scheme to form a set of FLOs. The $\rho_{i\sigma}$ in Eq.~(\ref{eq:pzsic}) are the square of the $i^{th}$ FLO. FLOs are used in both the PZSIC and LSIC calculations presented in this work. We note that in some early articles on FLOSIC, the term FLOSIC was occasionally used interchangeably with the term PZSIC. But other one electron SIC methods such as orbital scaling SIC (OSIC)\cite{vydrov2006scaling,vydrov2006simple,doi:10.1063/5.0004738} or LSIC can also be implemented using FLOs, and these may be referred to a FLOSIC implementation of OSIC or LSIC equations. In the FLOSIC approach, the optimal set of FLOs is obtained by minimizing the total energy with respect to the FOD positions using a gradient based algorithm. The expressions for the  FODs forces have been derived earlier for the PZSIC energy functional.\cite{doi:10.1063/1.4907592,PEDERSON2015153} In this work, we follow these earlier works to obtain the FOD force expression  by minimizing the LSIC energy  expression. 

\subsection{Local scaling Self-interaction-Correction method}\label{sec:localscaling}
Like the PZSIC method, the LSIC is also a one-electron SIC method in which the exchange-correlation energy is obtained as 
\begin{equation}\label{eq:LSIC-DFA}
  E_{XC}^{LSIC-DFA} = E_{XC}^{DFA}[\rho_{\uparrow},\rho_{\downarrow}]
                    - \sum_{i\sigma}^{occ}
  \left \{ U^{LSIC}[\rho_{i\sigma}]
    + E_{XC}^{LSIC}[\rho_{i\sigma},0] \right \}.
\end{equation}
Here, 
\begin{equation}\label{eq:lsiccoul}
  U^{LSIC}[\rho_{i\sigma}] = 
     \frac{1}{2} \int d\mathbf{r}\,
     z_{\sigma}(\mathbf{r}) \,
     \rho_{i\sigma}(\mathbf{r}) 
     \int d{\mathbf{r}\,}'\, \frac{\rho_{i\sigma}({\mathbf{r}\,}')}{\vert \mathbf{r}-{\mathbf{r}\,}'\vert}
\end{equation}
and
\begin{equation}\label{eq:lsicxc} 
  E_{XC}^{LSIC}[\rho_{i\sigma},0]
   =  \int d\mathbf{r}\,  z_{\sigma}(\mathbf{r})
   \varepsilon_{XC}^{DFA}([\rho_{i\sigma},0],\mathbf{r}). 
\end{equation}
$z_\sigma(\mathbf{r})$ is the local scaling factor or an iso-orbital indicator that is used to identify one-electron self-interaction regions. 
LSIC offers flexibility in that any suitable choice of iso-orbital indicator can be used to distinguish between many-electron and one-electron like regions.

In this work, we use  $z_\sigma(\mathbf{r})$  as the ratio of kinetic energy densities $\tau^W_\sigma(\mathbf{r})/\tau_\sigma(\mathbf{r})$. Total kinetic energy density $\tau_\sigma(\mathbf{r})$ and Weizs{\"a}cker kinetic energy density $\tau_\sigma^W(\mathbf{r})$ are defined as follows,\cite{weizsacker1935theorie}
\begin{equation}
  \tau_\sigma(\mathbf{r}) = \frac 1 2 \sum_i |\mathbf{\nabla} \psi_{i\sigma}(\mathbf{r})|^2,
\end{equation}
\begin{equation}
  \tau_\sigma^W(\mathbf{r}) = \frac{|\mathbf{\nabla} \rho_\sigma(\mathbf{r})|^2}{8 \rho_\sigma(\mathbf{r})}.
\end{equation}
$z_\sigma(\mathbf{r})$ approaches unity in the single electron regions and becomes vanishingly small in the uniform density regions. Thus LSIC corrects SIEs in the single electron region and reduces to DFA in uniform density regions recovering the uniform gas limit of parent semilocal DFA violated by the PZSIC method.\cite{doi:10.1063/1.5090534,doi:10.1063/1.5129533}

The functional derivative of $E^{LSIC}$ with respect to variation in density is written as follows,
\begin{align}
\label{eq:LSIC_functderiv}
  \frac{\delta E^{LSIC}}{\delta \rho(\mathbf{r})}  \Biggr|_{\rho=|\phi_{i\sigma}|^2}
    &=
    v^{DFA}(\mathbf{r})-
    \frac{1}{2}    \int \frac{  z_\sigma(\mathbf{r'})\rho_{i\sigma}(\mathbf{r'})}{|\mathbf{r}-\mathbf{r'}|} d\mathbf{r'} 
    - \frac{1}{2}    z_\sigma(\mathbf{r})\int \frac{  \rho_{i\sigma}(\mathbf{r'})}{|\mathbf{r}-\mathbf{r'}|} d\mathbf{r'}
    -z_\sigma(\mathbf{r})v_{xc}^{i\sigma}(\mathbf{r}) \nonumber\\
    & - \frac{1}{2} \sum_j^{N_\sigma}   \frac{\delta z_\sigma(\mathbf{r})}{\delta \rho}\rho_{j\sigma}(\mathbf{\mathbf{r}})\int \frac{\rho_{j\sigma}(\mathbf{r'})}{|\mathbf{r}-\mathbf{r'}|} d\mathbf{r'} 
    - \sum_j^{N_\sigma} \frac{\delta z_\sigma(\mathbf{r})}{\delta\rho} \epsilon_{xc}^{j\sigma}(\mathbf{r}) .
\end{align}
The ${\delta z(\mathbf{r})}/{\delta \rho}$ term can be evaluated following the approach by Neumann \textit{et al.}\cite{doi:10.1080/00268979600100011} The orbital dependent Fock matrix elements due to the last two terms in Eq.~(\ref{eq:LSIC_functderiv}) are obtained as follows,
\begin{align}
\label{eq:lsicmatelems}
  H_{\mu\nu}^{\sigma}
   = &-\sum_j^{N_\sigma}\int \Big(\varepsilon_{coul}^{j\sigma}(\mathbf{r})+\varepsilon_{xc}^{j\sigma}(\mathbf{r})\Big) \frac{\partial z_\sigma(\mathbf{r})}{\partial \rho(\mathbf{r})} \phi_\mu(\mathbf{r}) \phi_\nu(\mathbf{r}) d\mathbf{r} \nonumber\\
  &-\sum_j^{N_\sigma} \int \Big(\varepsilon_{coul}^{j\sigma}(\mathbf{r})+\varepsilon_{xc}^{j\sigma}(\mathbf{r})\Big) \frac{\partial z_\sigma(\mathbf{r})}{\partial \mathbf{\nabla}\rho(\mathbf{r})} \cdot \mathbf{\nabla}\Big(\phi_\mu(\mathbf{r}) \phi_\nu(\mathbf{r})\Big) d\mathbf{r} \nonumber\\
  &-\sum_j^{N_\sigma} \frac 1 2 \int \Big(\varepsilon_{coul}^{j\sigma}(\mathbf{r})+\varepsilon_{xc}^{j\sigma}(\mathbf{r})\Big) \frac{\partial z_\sigma(\mathbf{r})}{\partial \tau(\mathbf{r})} \mathbf{\nabla}\phi_\mu(\mathbf{r}) \cdot \mathbf{\nabla}\phi_\nu(\mathbf{r}) d\mathbf{r}.  
\end{align}
Here 
$\varepsilon_{coul}^{j\sigma}(\mathbf{r})$ and $\varepsilon_{xc}^{j\sigma}(\mathbf{r})$ are one-electron energy densities such that 
\begin{align}
  \varepsilon_{coul}^{j\sigma}(\mathbf{r}) &= \frac 1 2 \rho_{j\sigma}(\mathbf{r}) \int \frac{\rho_{j\sigma}(\mathbf{r'})}{|\mathbf{r}-\mathbf{r'}|} d\mathbf{r'} 
\end{align}
 and for the LSDA,
\begin{align}
  \varepsilon_{xc}^{j\sigma}(\mathbf{r}) &=  -\frac 3 4 \left(\frac 6 \pi \right)^{1/3}  \rho_{j\sigma}^{4/3}(\mathbf{r}) + \varepsilon_{c}\Bigl(r_s(\mathbf{r})=\left(\frac{3}{4\pi\rho_{j\sigma}(\mathbf{r})}\right)^{1/3},\zeta(\mathbf{r})=1\Bigl).
\end{align}

\subsubsection{Numerical Poisson solver}
Analytical evaluation of the second term in the right hand side of Eq.~(\ref{eq:LSIC_functderiv}) is not straightforward. We have therefore developed a numerical Poisson solver\cite{PHC22} based on the multi-center grid originally proposed by Becke.\cite{Becke1988d} A brief description of the implementation is as follows. First, a spherical mesh is constructed by multiplying Lebedev spherical mesh \cite{LEBEDEV197610} onto radial quadrature \cite{Mura1996b} for each atomic center. Due to the overlap between the spherical meshes from different atomic centers, a special type of integration weight function $w_{n}(\mathbf{r})$ \cite{becke1988e,Delley1990b,Stratmann1996b} that satisfies the condition $\sum_{n}w_{n}(\mathbf{r})=1$ is also constructed on each grid point to scale down the charge density as $\rho_{n}(\mathbf{r})=w_{n}(\mathbf{r})\rho(\mathbf{r})$ to eliminate double counting in the mesh overlapping region. As a result, the entire space is partitioned into small regions where each of these region confines a portion of charge $\rho_n$ and can be treated independently.

The advantage of this approach is that it allows each $\rho_n(\mathbf{r})$ to be expressed in spherical coordinate for efficient integration. By applying multipolar expansion on both charge distribution $\rho_{n}$ and its corresponding Coulomb potential $V_{n}$, the radial and the angular degrees of freedom become separable and the Poisson equation becomes a set of 1D differential equations for the radial parts of the potential $V_{lm}(r)$. The final full Coulomb potential due to $\rho_{n}(\mathbf{r})$ can be reconstructed on any given mesh using the following expression
\begin{align}
  V_{n}(\mathbf{r})=\sum_{l=0}\sum_{m=-l}^{l}V_{lm}(r)Y_{lm}(\theta,\phi) ,
\end{align} 
with interpolation of $V_{lm}(r)$ onto any given $r$.

To solve for the radial solutions $V_{lm}(r)$, common approaches are either to use a finite-difference method to solve the differential equation directly \cite{Becke1988d} or to integrate the Green's function (GF) solution to the Laplacian.\cite{Delley1990b} For the former approach, it uses the differential equations of the following form
\begin{align}
  \Big( 1 + p(r) \Big) \frac {\partial^2} {\partial r^2} U_{lm}(r) -q(r)U_{lm}(r)=f(r) 
\end{align}
with $V_{lm}(r)={U_{lm}(r)}/{r}$, and $p(r)$ and $q(r)$ are the coefficient functions where their values depend on the radial quadrature in use. For the later approach, the GF solution to Laplacian is written as
\begin{align}
  I_{lm}(r)=\frac{1}{r^{l+1}}\int_0^r r'^{l+2}\rho_{lm}(r')dr' + r^l \int_r^\infty \frac{\rho_{lm}(r')}{r'^{l-1}} dr'    
\end{align}
with $V_{lm}(r)=\frac{4\pi}{2l+1}I_{lm}(r)$.

Most of the latest implementations are based on GF approach \cite{Franchini2013b,Blum2009b} often combined with screening charge and Dunlap correction. \cite{dunlap1983fitting,Delley1990b,Franchini2013b,Blum2009b} In an orbital-by-orbital SIC calculation where the Coulomb potentials need to be evaluated for each occupied orbital, the technique mentioned above is performed repeatedly for each orbital density. By analyzing the sources of errors and accuracy in the two approaches, we put forth a hybrid method to optimize the radial solutions. More details of this hybrid approach can be found in Ref.~\onlinecite{PHC22}.

\subsection{Quasi-self-consistent LSIC}
A simplified {\it quasi}-self-consistent procedure can be obtained if one ignores the variation   of the iso-orbital (scaling factor) in Eq. (\ref{eq:LSIC_functderiv}) and replaces the $z_\sigma(\mathbf{r'})$ in the second term with $z_\sigma(\mathbf{r})$, which is mathematically valid only when $z_\sigma(\mathbf{r})$ is constant. Since the iso-orbital indicator $z_\sigma(\mathbf{r})$ varies substantially in space, the resulting Hamiltonian is not equivalent to Eq.~(\ref{eq:LSIC_functderiv}) except for one-electron systems. The resultant {\it quasi}-self-consistent Hamiltonian is given by
\begin{align}
  H_{i\sigma}^{quasi-LSIC}
   &=
   H^{DFA}
   - z_\sigma(\mathbf{r})\int \frac{  \rho_{i\sigma}(\mathbf{r'})}{|\mathbf{r}-\mathbf{r'}|} d\mathbf{r'}
   -z_\sigma(\mathbf{r})v_{xc}^{i\sigma}(\mathbf{r}) .
\end{align}
This Hamiltonian can be viewed as local scaling applied to orbital dependent SIC potential instead of SIC energy densities. As mentioned in the introduction, a few applications such as static dipole polarizabilities and dipole  moments of water clusters, polyacenes, etc. have been carried out using  {\it quasi}-self-consistent LSIC approach.\cite{waterpolarizability,doi:10.1063/5.0041265} Good results obtained in these studies shows that {\it quasi}-self-consistent LSIC approach can be useful to study some (especially density related) properties.

\subsection{FOD force and atomic force}
The optimal FLOs in the standard FLOSIC implementation are obtained by minimizing  the energy  with respect to FOD positions, in which determination of  Fermi orbital derivative terms are needed. The expression for the FOD force (energy derivative)\cite{doi:10.1063/1.4907592} can be simplified as follows 
\begin{align}
  \frac{dE^{SIC}}{da_m} &=\sum_{kl}\epsilon_{kl}^k \left \{\left\langle\frac{d\phi_k}{da_m}\bigg| \phi_l\right\rangle - \left\langle\frac{d\phi_l}{da_m}\bigg|\phi_k\right\rangle \right\},\\
  \epsilon_{kl}^k&=\langle\phi_l|V_k^{SIC}|\phi_k\rangle.
\end{align}
Since the term $\epsilon_{kl}^k$ here is the Lagrange multiplier matrix, we substitute this term with an equivalent terms constructed from Eq.~(\ref{eq:LSIC_functderiv}, \ref{eq:lsicmatelems}) for LSIC-LSDA. In our case, FOD optimizations are performed using this implementation. 

The Pulay atomic force\cite{PhysRevB.42.3276} for PZSIC-LSDA is given in Ref.~\onlinecite{doi:10.1002/jcc.25767} as follows,
\begin{align}
  \mathbf{F_\nu}^{Pulay}&=-2\sum_i^M \sum_{kl}^{N}c_k^i c_l^i \left\langle \frac{\partial \chi_k}{\partial \mathbf{R_\nu}} \bigg| H_i^{DFA-SIC}\bigg| \chi_l\right\rangle + 2 \sum_{ij}^M\Lambda_{ij}\sum_{kl}^{N}c_k^i c_l^j \left\langle \frac{\partial \chi_k}{\partial \mathbf{R_\nu}} \bigg| \chi_l \right\rangle, 
\end{align}
where $M$ is the number of occupied orbitals, $N$ is the size of the basis set, $\chi_k$ is the local basis function, $\Lambda_{ij}=\frac 1 2 (\lambda_{ij}-\lambda_{ji})$ is an element of the symmetrized Lagrange multiplier matrix, and $\mathbf{R}$ is the  nuclear position.
Here, to obtain atomic forces in the LSIC we modify the terms, $\langle \frac{\partial \chi_k}{\partial R_\nu} | H_i^{SIC}| \chi_l\rangle$ and $\Lambda_{ij}$, to accommodate the LSIC Hamiltonian. This modification allows simultaneous optimization of FODs and nuclei positions in principle, but in practice requires care as it adds more degrees of freedom to FOD optimization processes, and the FOD energy landscape is quite complicated. We have verified both the FOD forces and atomic forces by comparing them against the forces obtained numerically using finite difference methods.

\subsection{Computational details}
Both PZSIC and LSIC methods are implemented in the developmental version of the FLOSIC code.\cite{FLOSICcode,FLOSICcodep} FLOSIC code is based on the UTEP-NRLMOL code which itself is a modernized version of legacy NRLMOL (FORTRAN 77) code\cite{pederson2000strategies,PhysRevB.42.3276} with many additional new capabilities. We consider the LSDA functional since LSIC applied to LSDA is free from the gauge problem.\cite{doi:10.1063/5.0010375} 
Application of LSIC  to GGAs and particularly to SCAN meta-GGA provides little improvement compared to the PZSIC-SCAN except for the reaction barriers which are improved. The  cause for this shortcoming has been identified as a gauge problem as their exchange-correlation potentials and energy densities are not in the Hartree gauge and therefore require a gauge transformation\cite{doi:10.1063/5.0010375,PhysRevA.77.012509} or inclusion of a calibration function.\cite{doi:10.1063/1.4901238,C6CP00990E} For the LSIC-LSDA calculations in this work, the LSDA correlation functional parameterized as PW92\cite{PhysRevB.46.6671} is used. The  Gaussian basis sets of triple zeta quality\cite{PhysRevA.60.2840} are used. 
The basis sets used in this work are available at \url{https://github.com/FLOSIC/NRLMOL_basis_set}. The NRLMOL basis set is also accessible at Basis Set Exchange\cite{pritchard2019a} as DFO(+)-NRLMOL. The matrix elements are obtained numerically using variational mesh.\cite{PhysRevB.41.7453} Although the LSIC energy functional depends on the  kinetic energy densities, we have found that for the properties studied in this work, standard numerical grids used for SIC calculations are sufficiently accurate and no numerical instabilities were found. Generally, a kinetic energy density dependent (meta-GGA) functional requires denser mesh for numerical integration than needed for GGAs because the derivative terms arising from the iso-orbital indicator may oscillate abruptly in space. However, $z_\sigma(\mathbf{r})$ and its partial derivatives are smooth, and these quantities alone do not cause numerical instabilities. The self-consistent FLOSIC  calculations can be  performed either using optimized effective potential within the Krieger-Li-Iafrate approximation \cite{PhysRevA.103.042811} or  using the Jacobi update approach.\cite{PhysRevA.95.052505}
In this work we used the Jacobi update approach. The SCF convergence tolerance of $10^{-6} E_h$ for the total energy is used. For FOD optimizations, force criteria with $10^{-3} E_h/a_0$ are used. For all LSIC calculations, we used the converged densities and optimized FODs from PZSIC-LSDA as the starting point for perturbative, quasi-SCF, and full SCF calculations. Both potential mixing and Hamiltonian mixing can be used with LSIC-LSDA to accelerate the self-consistence convergence since the DFA Fock matrix does not depend on integration by parts and the Jacobi update approach applies the mixing algorithm to the DFA Fock matrix only.  

\section{Results}\label{sec:results}
In this section, we present the results for atomic total energies, atomization energies, barrier heights, HOMO eigenvalues, and equilibrium bond distances. 

\subsection{Atoms}
\subsubsection{Total Energy}
\begin{figure}[p]
\includegraphics[width=\columnwidth]{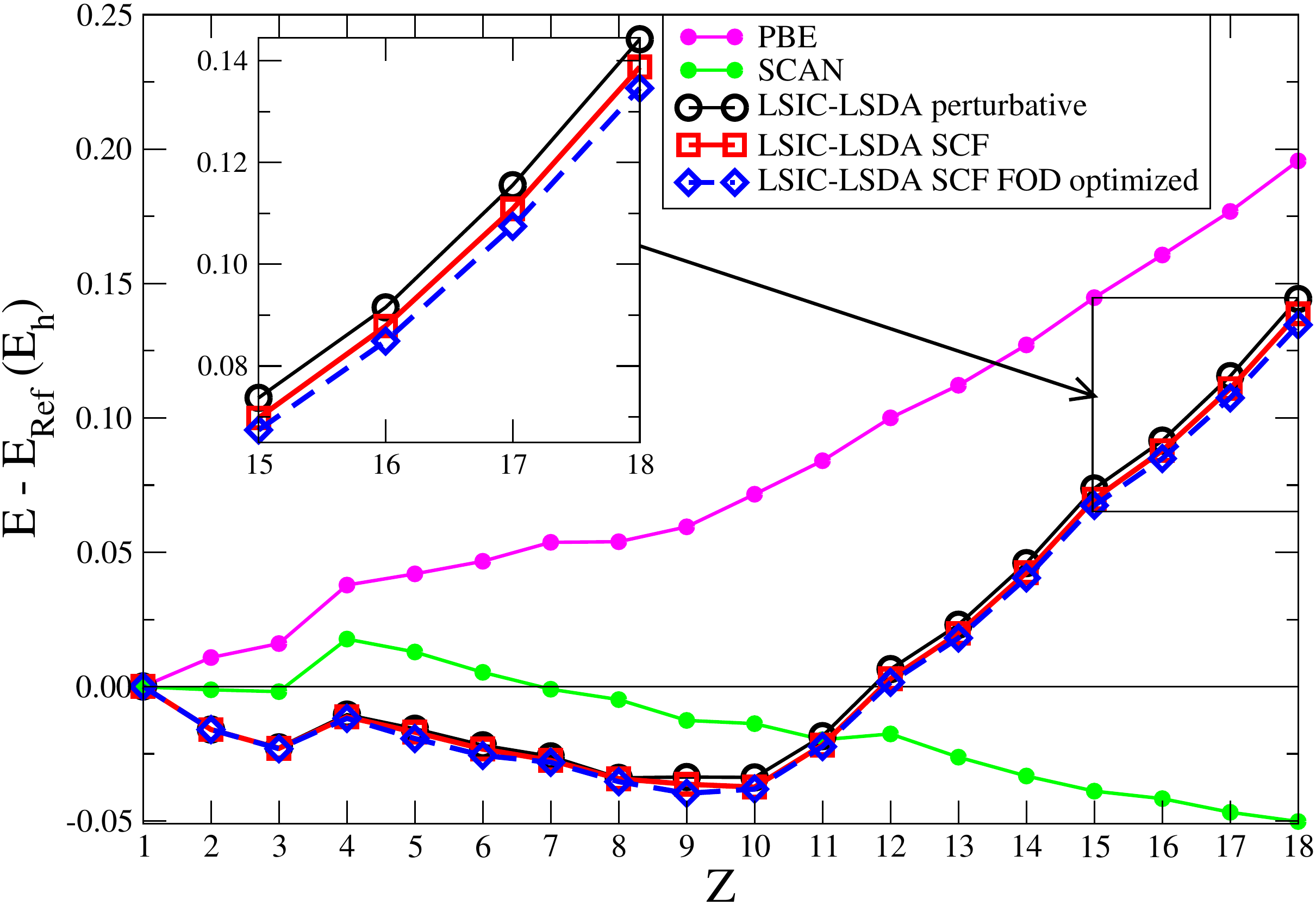}
\caption{\label{fig:atoms} Total energies of atoms (in $E_h$) compared against the reference values of Ref.~\onlinecite{PhysRevA.47.3649}.}
\end{figure}

We have calculated  self-consistent LSIC total energies of atoms from  hydrogen through argon. The deviations  of total energies from accurate reference values reported in Ref.~\onlinecite{PhysRevA.47.3649} are shown in Fig.~\ref{fig:atoms}. The mean absolute errors (MAE) of LSIC-LSDA method with respect to the reference values are summarized  in Table~\ref{tab:atoms}. We have also included the MAE of the perturbative and {\it quasi-}self-consistent LSIC-LSDA calculations to make the effect of full self-consistency  evident. As previously found, the perturbative LSIC-LSDA shows MAE of 0.041 $E_h$ and performs intermediate between the PBE and SCAN functionals. The self-consistency, on average,  lowers the energies of atoms by 2.4 $mE_h$. The optimization of the FODs further  lowers the atomic energies by 1.7 $mE_h$ on average. Due to these very small energy gains, the MAE of quasi-SCF, SCF, and SCF with FOD optimization are 0.040 $E_h$ in all three cases. These results show that self-consistency and the FOD optimization do not alter the performance of LSIC significantly for the atomic energies and that the perturbative LSIC-LSDA is sufficient to get good estimates of atomic energies.

\begin{table*}
\caption{\label{tab:atoms}MAE of atomic total energies (in $E_h$) with respect to the reference atomic energies from Ref.~\onlinecite{PhysRevA.47.3649}.}
\footnotetext{Reference \onlinecite{FLOSICSCANpaper}} 
\begin{ruledtabular}
\begin{tabular}{lc}
 Method & MAE ($E_h$) \\ \hline
LSDA$^\text{a}$	&0.726\\
PBE$^\text{a}$	&0.083\\
SCAN$^\text{a}$	&0.019\\
\hline
PZSIC-LSDA$^\text{a}$	&0.381\\
PZSIC-PBE$^\text{a}$	&0.159\\
PZSIC-SCAN$^\text{a}$	&0.147\\
\hline
LSIC-LSDA, perturbative	&0.041\\
LSIC-LSDA, quasi-SCF    &0.040\\
LSIC-LSDA, SCF          &0.040\\
LSIC-LSDA, SCF FOD optimized    & 0.040\\
\end{tabular}
\end{ruledtabular}
\end{table*}

\subsection{Atomization energies}
We studied the performance of LSIC on atomization energies using the AE6 set\cite{doi:10.1021/jp035287b} of molecules. AE6 set is  part of the Minnesota Database and is often used for benchmarking the performance of density functional approximations for atomization energies. The AE6 set is composed of six molecules: SiH$_4$, S$_2$, SiO, C$_3$H$_4$ (propyne), HCOCOH (glyoxal), and C$_4$H$_8$ (cyclobutane). The atomization energy $E_A$ is obtained as energy difference of the sum of fragment atomic energies $E^i_{fragment}$ and the complex energy $E_{complex}$ as follows
\begin{equation}
  E_A = \sum_i E^i_{fragment} - E_{complex}.
\end{equation}

The MAEs are summarized in Table~\ref{tab:atomization}. For the sake of comparison, we also included our previous results for atomization energies of two widely used semi-local functionals PBE-GGA and SCAN meta-GGA and their self-interaction-corrected counterparts. As previously shown, perturbative LSIC-LSDA with an MAE of 9.94 kcal/mol displays better performance than   PZSIC for three different kinds of functionals (PZSIC-LSDA, PZSIC-PBE, and PZSIC-SCAN). It is, to our knowledge, the first one-electron SIC scheme that gives atomization energies better than the PBE-GGA. LSIC-LSDA, however, falls short of bare SCAN functional. While there are meta-GGA and hybrid functionals (e.g., VSXC, B3LYP, and PBE0) that perform better than LSIC-LSDA for atomization energies, many of these functionals do not always provide good performance for barrier heights and dissociation energies.

The atomization energy of individual molecules in the AE6 set changes by up to 4 kcal/mol when self-consistency is introduced.  The MAE (9.53 kcal/mol), however,  decreases  very slightly by 0.41 kcal/mol compared to the perturbative LSIC. The self-consistency lowers molecular energies by about $2-6$ kcal/mol. This energy lowering is about the same order of magnitude as in case of atoms. Since atomization energies are obtained from the energy differences between atoms and complexes, the energy shift due to self-consistency is canceled out, and as a result, self-consistent LSIC performance remains close to the perturbative LSIC approach. The MAE difference between the two approaches is under 1 kcal/mol. 

We repeated the atomization energy calculations for the LSIC using the quasi-SCF approach. We find that quasi-SCF shows smaller MAE of 6.57 kcal/mol. In quasi-SCF, energy shifts with respect to the perturbative LSIC energies are positive, and larger molecules tend to experience larger energy shifts (+32.1 kcal/mol for cyclobutane) than the smaller molecules (+5.5 kcal/mol for SiO), resulting thereby in decrease in the MAE of atomization energies. Finally, relaxation of the FODs results in further decrease in MAE of self-consistent LSIC by roughly 1 kcal/mol (MAE 8.66 kcal/mol). Overall performance remained unchanged from the perturbative approach to SCF to FOD optimized SCF. To summarize the results of this section, the performance of self-consistent LSIC with and without FOD optimization remains similar to that of perturbative LSIC that uses PZSIC orbital densities.

\begin{table*}
\caption{\label{tab:atomization}MAE of AE6 set of atomization energies.}
\footnotetext{Reference \onlinecite{doi:10.1063/1.5129533}}
\footnotetext{Reference \onlinecite{doi:10.1063/5.0004738}} 
\footnotetext{Reference \onlinecite{B316260E}}
\begin{ruledtabular}
\begin{tabular}{lc}
 Method & MAE (kcal/mol) \\ \hline
LSDA$^\text{a}$	&74.26\\
PBE$^\text{a}$	&13.43\\
SCAN$^\text{a}$	&2.85\\
\hline
PZSIC-LSDA$^\text{a}$	&57.97 \\
PZSIC-PBE$^\text{a}$	&18.83\\
PZSIC-SCAN$^\text{b}$	&26.52\\
\hline
LSIC-LSDA, perturbative	&9.94\\
LSIC-LSDA, quasi-SCF & 6.57 \\
LSIC-LSDA, SCF   &9.53\\
LSIC-LSDA, SCF FOD optimized    & 8.66 \\
\hline
B3LYP/MG3S$^\text{c}$ & 3.2 \\
PBE0/MG3S$^\text{c}$ & 5.4\\
\end{tabular}
\end{ruledtabular}
\end{table*}

\subsection{Barrier heights}
We used the BH6 set\cite{doi:10.1021/jp035287b} of reactions to study the LSIC performance on barrier heights. BH6 is a set of three hydrogen transfer reactions: (1) OH + CH$_4$ $\rightarrow$ CH$_3$ + H$_2$O, (2) OH + H $\rightarrow$ O + H$_2$, and (3) H + H$_2$S $\rightarrow$ HS + H$_2$. For each reaction, single point energies of left and right hand side and saddle point of the reactions were calculated, and then forward and reverse barrier heights were computed. Simultaneous accurate description of barrier heights and atomization is challenging not only for uncorrected DFAs but also for self-interaction correction methods.\cite{Janesko2008} Improvements in barrier  heights  usually occurs at the expense of accuracy of atomization energies. Our previous results of perturbative LSIC showed that it goes beyond this paradoxical  behavior of well-known PZSIC method and can provide good description of both barrier heights and atomization energies. The performance of LSIC method(s) is summarized in Table~\ref{tab:barrierheights}. All three DFAs in the table underestimate barrier heights where the transition state energies are predicted too low due to SIE in the energy functional. As such, LSDA show MAE of 17.6 kcal/mol. All PZSIC calculations improve the barrier heights for semilocal functionals with MAE ranging from 3.0 kcal/mol of PZSIC-SCAN to 4.9 kcal/mol of PZSIC-LSDA. Application of LSIC-LSDA raises the barriers and further reduces MAE down to 1.3 kcal/mol. We note that LSIC barrier heights are not necessarily between those predicted by the bare DFA and PZSIC-DFA since energy shifts vary for reactant, product, and transition states. The MAE of self-consistent LSIC is 1.1 kcal/mol. Once again we see that full self-consistency  performs similar to perturbative LSIC (MAE 1.3 kcal/mol). This holds true even after FODs are optimized (MAE 1.2 kcal/mol). Although quasi-SCF also gives smaller MAE of 1.5 kcal/mol, deviation in barrier heights from molecule to molecule is far more compared to perturbative LSIC and self-consist LSIC. It is likely that the performance of quasi-SCF would deviate from the others for more diverse data set. For a larger BH76 dataset, perturbative LSIC showed MAE of 3.7 kcal/mol.\cite{doi:10.1063/5.0070893} 

\begin{table*}
\caption{\label{tab:barrierheights}MAE of BH6 set of barrier heights.}
\footnotetext{Reference \onlinecite{doi:10.1063/1.5129533}}
\footnotetext{Reference\onlinecite{doi:10.1063/5.0010375}}
\footnotetext{Reference \onlinecite{B316260E}}
\footnotetext{Reference \onlinecite{doi:10.1021/jp035287b}}
\begin{ruledtabular}
\begin{tabular}{lc}
Method & MAE (kcal/mol) \\ \hline
LSDA$^\text{a}$	& 17.6\\
PBE$^\text{b}$	& 9.6\\
SCAN$^\text{a}$	& 7.9\\
\hline
PZSIC-LSDA$^\text{a}$	& 4.9\\
PZSIC-PBE$^\text{a}$	& 4.2\\
PZSIC-SCAN$^\text{a}$	& 3.0\\
\hline
LSIC-LSDA, perturbative$^\text{a}$	& 1.3\\
LSIC-LSDA, quasi-SCF & 1.5 \\
LSIC-LSDA, SCF   & 1.1 \\
LSIC-LSDA, SCF FOD optimized    & 1.2\\
\hline
B3LYP/MG3S$^\text{c}$ & 4.7\\
PBE0/MG3S$^\text{c}$ & 4.6\\
HF/MG3S$^\text{d}$ & 12.3\\
\end{tabular}
\end{ruledtabular}
\end{table*}

\subsection{HOMO eigenvalues}
\begin{figure}[p]
\includegraphics[width=\columnwidth]{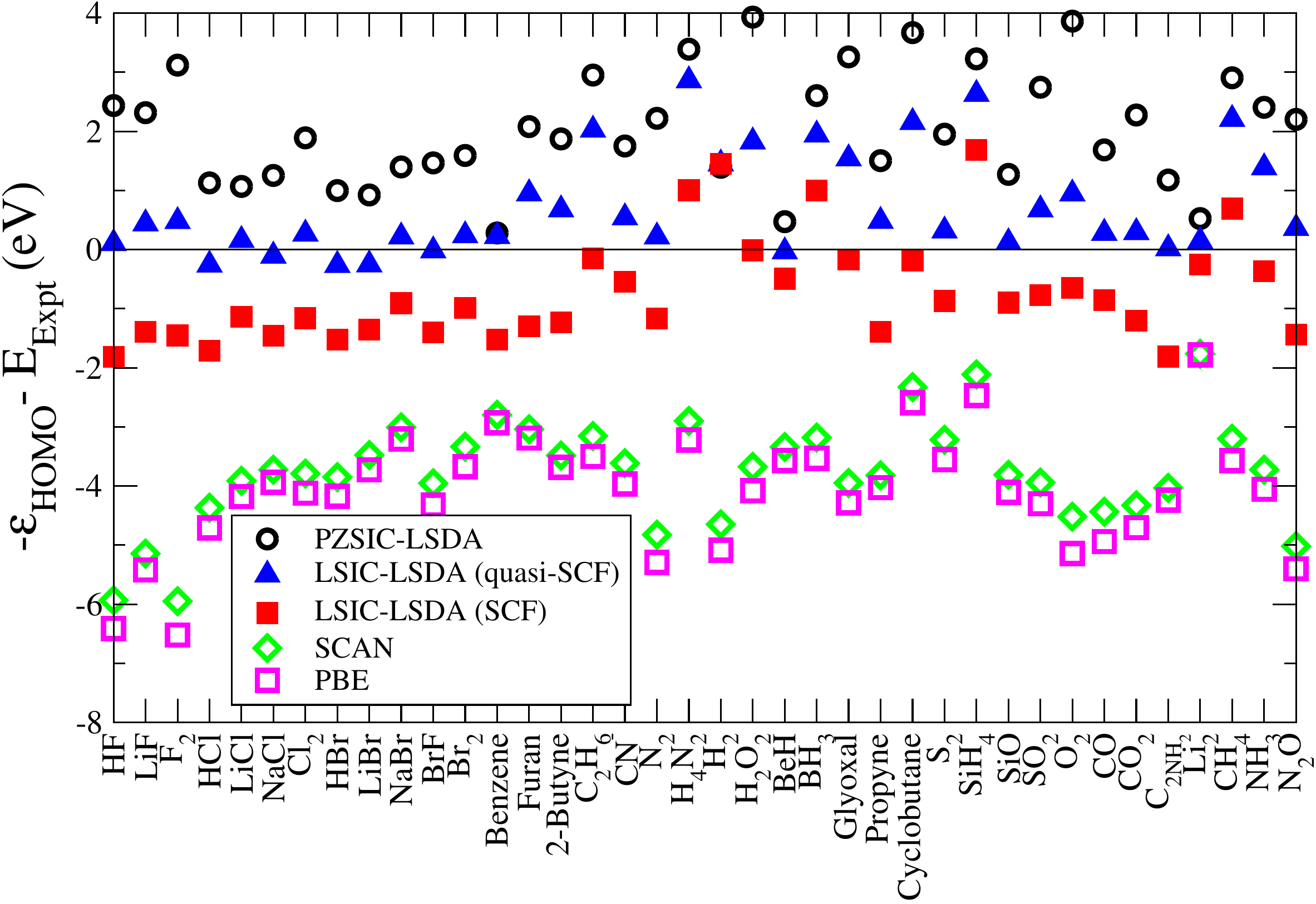}
\caption{The deviations of $-\varepsilon_{HOMO}$ against the experimental IP (in eV).
}\label{fig:ehomo}
\end{figure}

In DFT, the negative of the HOMO eigenvalue equals the first ionization potential.\cite{perdew1982density,PhysRevA.30.2745,perdew1997comment,PhysRevB.60.4545} The validity of this relation in the approximate density functional calculations depends on the quality of the asymptotic description of the effective potential which is affected by the SIE. In most DFAs, the absolute of the HOMO eigenvalue underestimates the first ionization potential substantially due to the shallow asymptotic nature of the approximate exchange potential. Correcting for the long-range description of DFA exchange potential improves the accuracy of HOMO\cite{doi:10.1021/acs.jctc.5b00873} and can result in  bound atomic anions.\cite{PhysRevLett.94.043002} PZSIC can improve the HOMO eigenvalues over those from DFAs by improving the asymptotic description of the exchange potential. Earlier perturbative LSIC calculations could not assess the quality of (negative of) HOMO eigenvalues in approximating the first ionization energies. We compare the PZSIC and LSIC  HOMO eigenvalues of a set of molecules and compare them against the experimental ionization energies reported in the NIST database. We present the deviations in Fig.~\ref{fig:ehomo}. As expected, the semilocal functionals PBE and SCAN underestimate the ionization energies with  MAE of 4.02 eV and  3.70 eV, respectively. On the other hand, PZSIC-LSDA overestimates with MAE of 2.10 eV where the majority of the values deviate between 0--4 eV above the experimental values. The valence electrons in the PZSIC are too strongly bound. We find that LSIC HOMO eigenvalues fall between the DFA and PZSIC HOMO eigenvalues. LSIC  HOMO eigenvalues exhibit a trend that is opposite to the PZSIC eigenvalues. They are  underestimated with a MAE 1.04 eV, which is roughly half the MAE of PZSIC-LSDA. Interestingly, quasi-SCF LSIC shows even better agreement with experiment (MAE 0.77 eV) than the full SCF case. As  quasi-LSIC can be viewed as local scaling applied to the the potential instead of energy, this suggests that for some properties scaling the potential rather than the energy density can be beneficial.

\subsection{Bond lengths}
\begin{figure}
  \centering
  \includegraphics[width=\columnwidth]{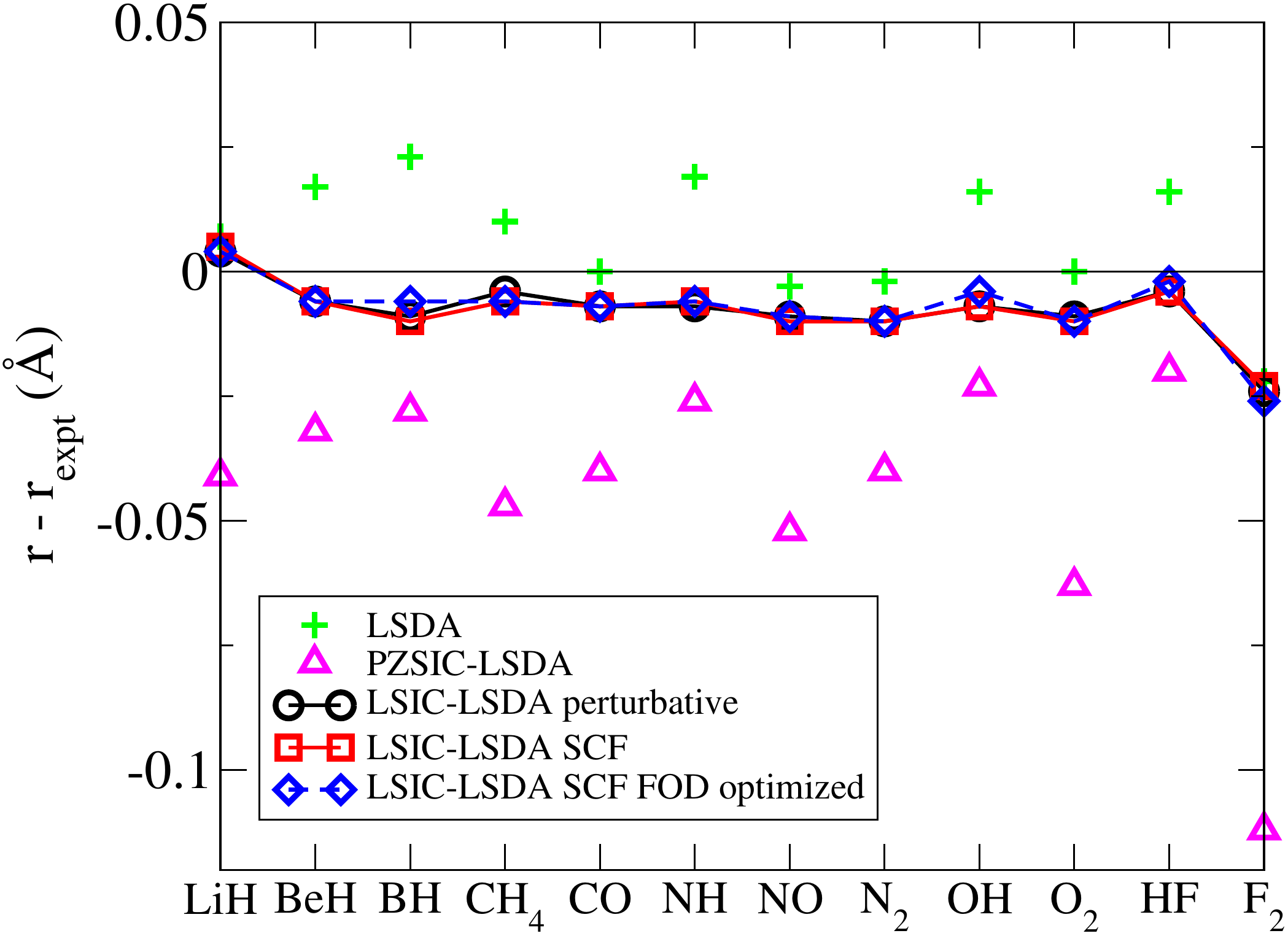} 
  \caption{Equilibrium bond distances (in {\AA}) compared against the experimental values from Ref. \onlinecite{lide2004crc}.}
  \label{fig:bondlen}
\end{figure}

We investigated how well LSIC performs for bond distances. The uncorrected LSDA performs fairly well in predicting bond distances. On the other hand, the PZSIC is known to predict bond lengths of molecules that are too short in comparison to the experiments.\cite{goedecker1997critical,doi:10.1063/1.5125205,vydrov2006scaling,csonka1998inclusion} For this reason, PZSIC is not suited for determining an optimal geometry, and in many cases PZSIC calculations are commonly performed on geometries obtained using  PBE, PBE0, or B3LYP functionals or using geometries from beyond Hartree-Fock methods. Vydrov and Scuseria investigated\cite{vydrov2006simple} the equilibrium bond distances of 12 small molecules and found mean errors of -0.045 \AA{} for PZSIC-LSDA and -0.024 \AA{} for PZSIC-PBE against experimentally found values. Those 12 molecules are LiH, BeH, BH, CH$_4$, CO, NH, NO, N$_2$, OH, O$_2$, HF, and F$_2$. This short bond length behavior is partly because PZSIC provides excessive corrections along bonding regions. Here, we examined how well LSIC performs in  determining bond lengths using the same set of 12 molecules used in the Vydrov and Scuseria's study. In the FLOSIC formalism, the geometry optimization and the FOD optimization can be, in principle, performed simultaneously.\cite{doi:10.1002/jcc.25767} However, our experience is that the FLOs (and FODs) are very sensitive to small changes in the geometry, which makes simultaneous optimization of FODs and atomic positions difficult. To obtain the equilibrium bond distances, we perform PZSIC and LSIC calculations at five geometries around the experimental bond lengths. This range of geometries covers minima in all three DFA, PZSIC, and LSIC cases. We subsequently used  the fitting function $f(x) = a + b(x-c)^2 + d(x-c)^3$ to determine the equilibrium bond lengths. During the fitting, the cubic term was introduced to reduce the fitting errors, and the parameter $d$ was initially set to a very small value. The comparisons against the experimental values reported in Ref.~\onlinecite{lide2004crc} were made and shown in Fig.~\ref{fig:bondlen}. We report MAE of each method. LSDA shows reasonable estimate with MAE of 0.011 {\AA} whereas PZSIC-LSDA shows apparent underestimation (MAE 0.044 {\AA}). The LSIC bond lengths (MAE 0.008 {\AA}) in comparison to PZSIC-LSDA are longer and are in better agreement with experimental bond lengths than the PZSIC-LSDA and LSDA for this set of molecules. In most cases, SCF and FOD optimization have small effect on the bond length. However, optimizing FODs tends to slightly improve bond lengths especially for BH and OH.
    
\begin{figure}
  \centering
  \includegraphics[width=\columnwidth]{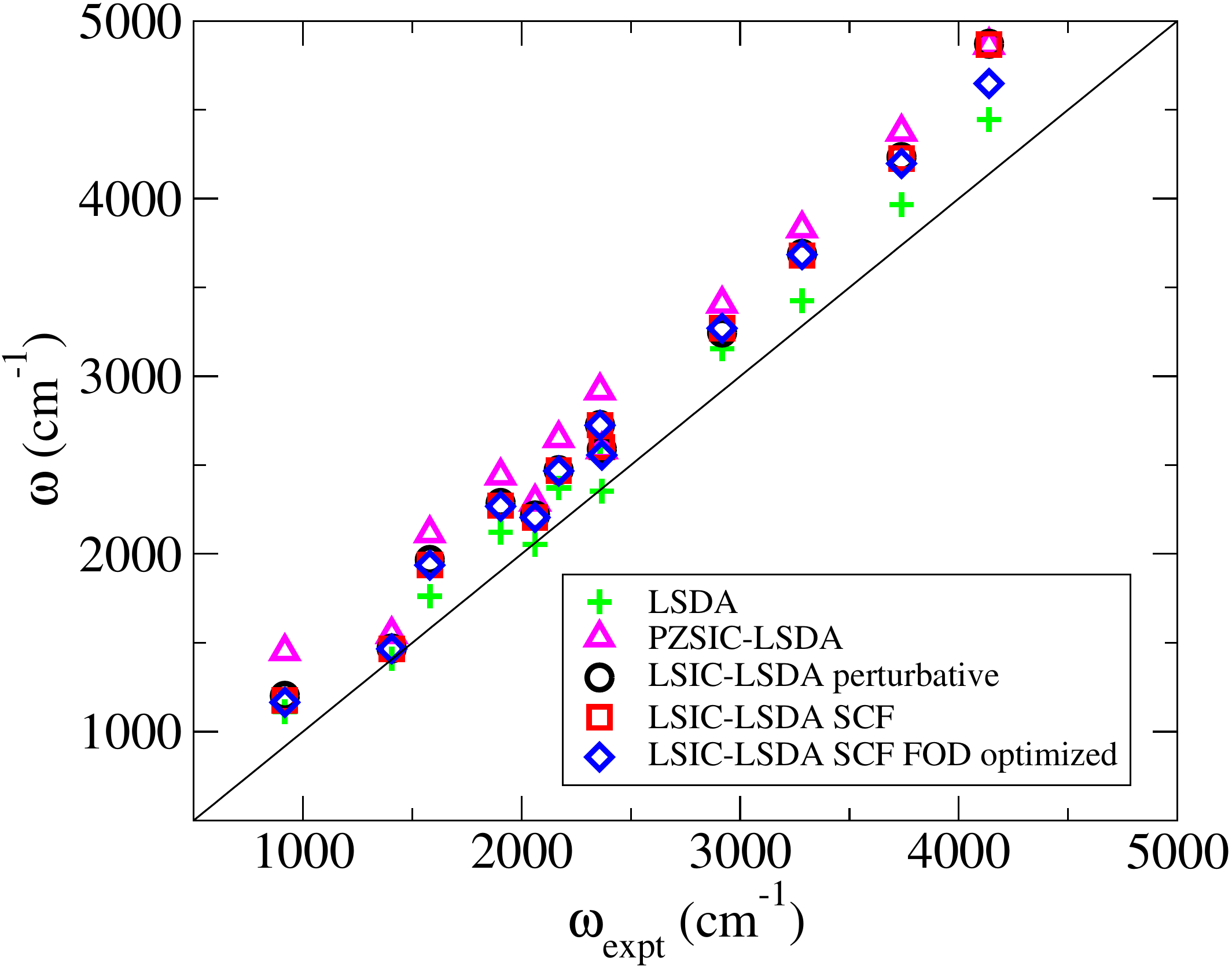} 
  \caption{Harmonic frequency obtained from the fitting function (in cm$^{-1}$) compared against the experimental values from Refs. \onlinecite{doi:10.1063/1.2436891,doi:10.1063/1.555560}.}
  \label{fig:harmonicf}
\end{figure}

Additionally, the parameter $b$ used in the fitting function can be used to estimate harmonic frequency $\omega$ of the set of diatomic molecules. The values of $\omega$ obtained from LSDA  show better agreement with experiment than both the PZSIC and LSIC values (cf. Fig.~\ref{fig:harmonicf}). PZSIC frequencies $\omega$ are about 13\% higher than the LSDA. The chemical bonds in PZSIC are {\em vibrationally blueshifted} compared to bonds in the LSDA. As seen for many other properties, LSIC  corrects the PZSIC frequencies and predicts $\omega$ values that are intermediate between the LSDA and PZSIC.

\section{Conclusion}
To summarize, in this work we present the self-consistent implementation of the LSIC method. We have presented the pertinent equations and details of the code implementation using the integration by part approach. The performance of self-consistent LSIC method for atomic energies, atomization energies, and reaction barrier heights  is assessed using standard benchmark databases. The results are compared with PZSIC-LSDA, the perturbative one-shot LSIC using PZSIC densities, as well as  the  {\it quasi}-self-consistent approach. In most cases, the {\it quasi}-self-consistent approach provides results comparable to  the full self-consistent LSIC method. In addition, we have obtained the ionization  energies from the  HOMO eigenvalues and the bond lengths of a set of molecules with the full self-consistent LSIC method. The HOMO eigenvalues of LSIC fall between those of LSDA and PZSIC-LSDA. Interestingly, a simpler quasi-SCF LSIC provides better agreements between HOMO eigenvalues and experimental ionization energies indicating that LSIC applied to the potential may be more useful approach for certain properties. Finally, we investigated equilibrium bond distances of dimers. PZSIC-LSDA tend to shorten the bond lengths which is corrected by  the LSIC. The bond distances of a chosen set of molecules obtained with LSIC method are somewhat shorter than those predicted by the LSDA but are in slightly better agreement with experimental values. From perturbative LSIC to self-consistent LSIC, energy as a function of bond distance is a uniform shift, and there is no significant change in the estimated bond lengths. The present results show that  perturbative LSIC on the top of PZSIC  and the self-consistent LSIC approaches perform similarly for the properties considered in this work. The cost of perturbative LSIC is insignificant if PZSIC densities are available. The self-consistent results of this work confirms our  previous conclusions that the LSIC method provides superior results than the well-known PZSIC method for all properties studied here thus providing an attractive approach to eliminate  self-interaction-errors that pervade most density functional studies. Further improvements in the performance of LSIC method may be possible by designing better iso-orbital indicators that work with LSIC or by designing the more sophisticated  density functionals for the LSIC method than the simple LSDA functional  considered in this work. The  vdW and nonlocal corrections can be also included to LSIC-DFA to improve its performance for the weakly bonded systems. The self-consistent implementation of LSIC method presented in this work also opens up its applications beyond energetic properties. Such studies will be carried out in future. 

The results of this work and previous one shot LSIC results show that accurate description of several electronic properties can be obtained from the simplest local spin density functional by removing the self-interaction errors using appropriately designed self-interaction correction method. The one-electron SIC method(s) is often considered  synonymous with PZSIC method and the success and failures of PZSIC have been often misinterpreted as those of one-electron SIC method(s). We hope that the present LSIC results along with those in earlier works with perturbative or quasi-self-consistent LSIC method help remove this general misconception. More studies, especially in cases where the effect of SIE are pronounced such as the transition metals, lanthanides and actinides complexes, or ions in solution are needed to obtain comprehensive picture of the scope of LSIC method.

\section*{Data Availability Statement}
The data that supports findings in this article is provided within the manuscript and supplementary information.

\begin{acknowledgments}
Authors dedicate this work to Dr. Brett Dunlap on his 75th birthday. Authors acknowledge discussions with Drs. Carlos Diaz and Luis Basurto. This work was supported by the US Department of Energy, Office of Science, Office of Basic Energy Sciences, as part of the Computational Chemical Sciences Program under Award No. DE-SC0018331. Support for computational time at the Texas Advanced  Computing Center (TACC), the Extreme Science and Engineering Discovery Environment (XSEDE), and National Energy Research Scientific Computing Center (NERSC) is gratefully acknowledged.  
\end{acknowledgments}

\bibliography{refs}

\end{document}